\documentclass[sigconf, nonacm]{acmart}
\AtBeginDocument{%
  \providecommand\BibTeX{{%
    \normalfont B\kern-0.5em{\scshape i\kern-0.25em b}\kern-0.8em\TeX}}}

\copyrightyear{2024}
\acmYear{2024}
\setcopyright{rightsretained}
\acmConference[CF '24]{21st ACM International Conference on Computing Frontiers}{May 7--9, 2024}{Ischia, Italy}
\acmBooktitle{21st ACM International Conference on Computing Frontiers (CF '24), May 7--9, 2024, Ischia, Italy}\acmDOI{10.1145/3649153.3649182}
\acmISBN{979-8-4007-0597-7/24/05}
\begin{document}

\title{Beyond the Buzz: Strategic Paths for Enabling Useful NISQ Applications}

\author{Pratibha Raghupati Hegde$^1$, Oleksandr Kyriienko$^2$, Hermanni Heimonen$^3$, Panagiotis Tolias$^4$, Gilbert Netzer$^1$, Panagiotis Barkoutsos$^5$, Ricardo Vinuesa$^6$ $^7$, Ivy Peng$^1$, Stefano Markidis$^1$}
\affiliation{%
\institution{$^1$KTH Royal Institute of Technology, Stockholm, Sweden;\\$^2$Department of Physics and Astronomy, University of Exeter, Exeter, United Kingdom;\\$^3$IQM Quantum Computers, Espoo, Finland; \\$^4$Space and Plasma Physics, KTH Royal Institute of Technology, Stockholm, Sweden;\\$^5$PASQAL SAS, Paris, France;\\$^6$FLOW, Engineering Mechanics, KTH Royal Institute of Technology, Stockholm, Sweden;\\$^7$Swedish e-science Research Centre (SeRC), Stockholm, Sweden
\country{}
}
}

\renewcommand{\shortauthors}{}

\begin{abstract}
  There is much debate on whether quantum computing on current NISQ devices, consisting of noisy hundred qubits and requiring a non-negligible usage of classical computing as part of the algorithms, has utility and will ever offer advantages for scientific and industrial applications with respect to traditional computing. In this position paper, we argue that while real-world NISQ quantum applications have yet to surpass their classical counterparts, strategic approaches can be used to facilitate advancements in both industrial and scientific applications. We have identified three key strategies to guide NISQ computing towards practical and useful implementations. Firstly, prioritizing the identification of a "killer app" is a key point. An application demonstrating the distinctive capabilities of NISQ devices can catalyze broader development. We suggest focusing on applications that are inherently quantum, e.g., pointing towards quantum chemistry and material science as promising domains. These fields hold the potential to exhibit benefits, setting benchmarks for other applications to follow. Secondly, integrating AI and deep-learning methods into NISQ computing is a promising approach. Examples such as quantum Physics-Informed Neural Networks and Differentiable Quantum Circuits (DQC) demonstrate the synergy between quantum computing and AI. Lastly, recognizing the interdisciplinary nature of NISQ computing, we advocate for a co-design approach. Achieving synergy between classical and quantum computing necessitates an effort in co-designing quantum applications, algorithms, and programming environments, and the integration of HPC with quantum hardware. The interoperability of these components is crucial for enabling the full potential of NISQ computing. In conclusion, through the usage of these three approaches, we argue that NISQ computing can surpass current limitations and evolve into a valuable tool for scientific and industrial applications. This requires an approach that integrates domain-specific killer apps, harnesses the power of quantum-enhanced AI, and embraces a collaborative co-design methodology. 
  
\end{abstract}




\begin{teaserfigure}
\centering
  \includegraphics[width=0.8\textwidth]{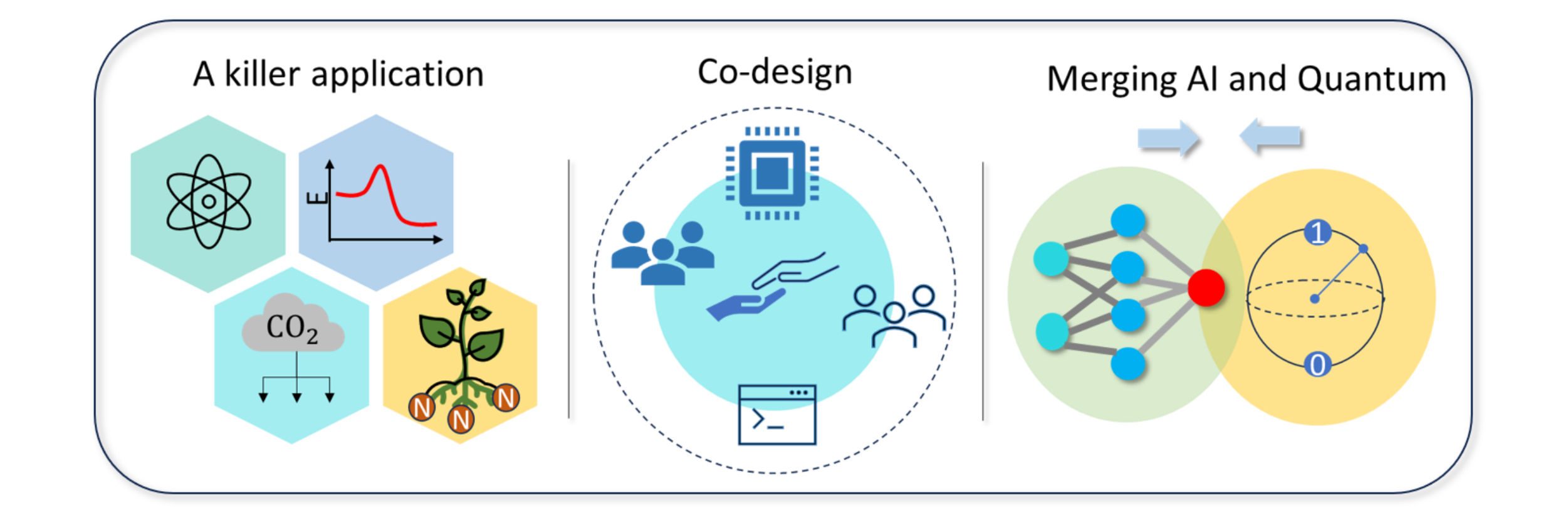}
\caption{Strategic pathways towards realizing a practical quantum advantage using NISQ devices}
\Description{}
\label{fig:teaser}
\end{teaserfigure}


\maketitle

\section{Introduction}

    On the path to fault-tolerant quantum computers that will be built with thousands/million of logical qubits, intermediate-scale quantum systems with up to hundreds of qubits serve as valuable experimental settings. For instance, they help to improve the quantum computing technologies and devise software and algorithms in preparation for full-fledged error-corrected quantum computing systems. Examples of such systems are Google’s Sycamore and IBM’s Osprey processors. The practical implementation of such systems has proved to be a challenging task as quantum systems are highly prone to decoherence and a quantum circuit has to be executed within the coherence time of qubits. Furthermore, the gates are imperfect leading to inevitable errors, and we need effective error correction protocols to counteract. These quantum computers belong to the Noisy Intermediate Scale Quantum (NISQ) era, a term which was coined by J. Preskill~\cite{Preskill2018:NISQ}. 

    NISQ systems are and have been very valuable experiments to further understanding and improving the fundamental technologies and the building blocks of quantum computing systems. One of the major and most important consequences of making NISQ systems open and available to the enthusiastic research community is the evolution of an ecosystem of software tools that allow us to develop and deploy the first quantum codes to tackle problems with scientific and industrial relevance. Differently from established quantum algorithms, such as Shor’s and Grover’s algorithms, NISQ algorithms~\cite{bharti2022noisy} and applications need to cope with noisy qubits as well as to include a non-negligible and critical part running on classical computing. Because of these constraints and limitations, an intense debate on the usefulness of NISQ applications and algorithms arose, and whether or not there will be any advantage in using NISQ applications when compared to classical computing~\cite{lau_nisq_2022, hoefler:onqadvantage}.

    Our position is that NISQ usefulness and advantage can be only achieved via a strategic approach towards the development of NISQ applications. In this position paper, we discuss the important needs of the NISQ era quantum computing which can drive them towards practical applications:
    \begin{enumerate}
    \item The quest for a killer application of NISQ devices which would further motivate the exclusive adoption of NISQ computing in different fields of science and industry.
    \item The combination of artificial intelligence with quantum algorithms for useful applications.
    \item The need for co-design of quantum hardware and software, a task that requires a highly interdisciplinary team of researchers.
    \end{enumerate}

\section{On the quest for the killer app}

The identification of a NISQ scientific or industrial application that demonstrates quantum usefulness and advantage, what we call a killer app, can drive a widespread adoption of the technique and cross-fertilize different fields. We can argue that cryptography and Shor's-type algorithms constitute the main application area for large-scale fault-tolerant quantum computing. However, it is challenging to use NISQ devices to implement theoretically successful algorithms for useful applications, since they typically require deep circuits with many qubits to solve difficult problems and hence to demonstrate the power of quantum computing~\cite{Gidney2021howtofactorbit}. 

This leaves us with the necessity to find the killer application of NISQ devices that is compelling and groundbreaking enough to drive widespread adoption of a particular technology or platform. Variational quantum algorithms (VQAs) constitute a promising avenue to utilize NISQ devices for useful applications. They are hybrid (classical-quantum) approaches which typically use a shallow depth parametrized quantum circuit ansatz where parameter optimization is done on classical computers by minimizing a cost function. In addition, this approach can yield valuable results even in the presence of noise. Given these features, the killer application for NISQ systems could be one which can be solved using VQAs. 
Here we briefly elucidate some of the NISQ applications.

\noindent \textbf{Quantum simulations.} As famously suggested by Richard Feynman at the dawn of quantum computing, quantum computers are a natural fit for simulating quantum systems~\cite{daley_practical_2022}. Variational Quantum Eigensolver (VQE) is a variational technique to find the ground state of a molecule by minimizing the energy with respect to variational parameters~\cite{peruzzo_variational_2014}. This algorithm is particularly useful in quantum chemistry applications such as the calculation of molecular spectral properties~\cite{colless:quantchem}, the determination of the optimal geometry of molecules~\cite{delgado:quantchem}, and the simulation of chemical reactions~\cite{motta:quantchem}. An example is the problem of finding an energy-efficient catalyst which accelerates nitrogen fixation. An approach to simulate this computationally expensive problem on the NISQ computer was shown in~\cite{reiher:nitrogen_fixation}. Another example is that of carbon dioxide conversion into value added safely storable chemicals that could even nullify such anthropogenic greenhouse gas emissions~\cite{greene-diniz_modelling_2022}.

\noindent \textbf{Quantum optimization.} Combinatorial optimization problems have many real-world applications such as financial portfolio optimization, drug discovery, traffic flow optimization, and logistics. Quantum Approximate Optimization Algorithm (QAOA) is a valuable tool in the NISQ era to obtain approximate solutions to such problems~\cite{farhi:QAOA}. This is a variational quantum algorithm that implements a sequence of quantum circuits with parameterized unitary evolution corresponding to the adiabatic state preparation. Quantum annealing (QA) is another quantum adiabatic algorithm that exploits quantum fluctuations to drive the system from an easy to prepare ground state of a simple Hamiltonian to the ground state of a more complicated problem Hamiltonian~\cite{kadowaki:qa}. Both QAOA and QA have so far been able to exhibit only polynomial speedup which is an impressive feat for small scale problems. However, it will be truly revolutionary if a killer application is found in the field of optimization, where industrial impact is tremendous. 

On the quest of the NISQ killer application, we argue that application simulations with an inherent quantum nature will unlock the full power of quantum computing. These are algorithms and applications that directly and natively rely on superposition and entanglement to formulate a problem to solve, e.g., quantum simulations. Examples of such applications are quantum chemistry, material science, protein folding problems~\cite{robert_resource-efficient_2021} and fully or partially degenerate plasma applications as those arising in High Energy Density Physics (HEDP)~\cite{Hatfield_HEDP_2021} and Warm Dense Matter (WDM)~\cite{Dornheim_WDM_2023}. Another set of algorithms that might benefit are the ones that have governing equations that match or have strong similarities with quantum mechanics simulations or methods for solving quantum equations. An example concerns equations for the modeling of fluid turbulence that rely on quantifying correlations between different length scales using methods inspired by quantum many-body physics.

\section{AI \& Quantum}

Artificial intelligence (AI) has taken almost every field of industry by storm and rigorous adoption in science is taking place as well in the recent years. Combining quantum computing techniques with AI is beyond the mere combination of two popular technologies, since it lays a path towards quantum advantage. With the property of quantum superposition, parallelism is native to quantum systems. Further, the presence of quantum entanglement among qubits, which is stronger than classical correlation, should be able to recognize complex patterns in data that are beyond the reach of classical machine learning algorithms~\cite{rebentrost:qsvm}. This intuition has driven the emergence of quantum machine learning (QML), where models are built with the help of quantum computers~\cite{biamonte:qml, harrow:hhl, aimeur:q-unsupervised, QPCA:lloyd, wiebe:qnearest-neighbors}. QML techniques can already be developed and tested on NISQ devices as variational quantum circuits are operationally similar to deep neural networks. They can be trained by minimizing a cost function and therefore QML approaches suit NISQ devices. Together with the enhanced "expressivity" of quantum systems, QML techniques indeed have the potential to surpass classical machine learning.  
The representation of classical data as a quantum mechanical problem is one of the major challenges of QML techniques creating speculations about their applicability for real world problems. However, we believe that QML is certain to show speedup in processing quantum data. Such quantum data could be the outcome of quantum sensing experiments, which are making promising developments towards precise measurements. QML will be a subroutine which further enhances the efficiency of such quantum metrology techniques. 

Combination of AI and quantum computing on NISQ devices can be adopted for scientific machine learning applications such as solving differential equations~\cite{lubasch:vqa}. 
A recently developed NISQ algorithm for solving differential equations is Differentiable Quantum Circuits (DQC)~\cite{kyriienko:dqc} as Universal Function Aprroximators (UFAs). This algorithm is inspired by Physics Informed Neural Networks (PINNs) which are beneficial when there is availability of some data of the physical system and the underlying equation of motion is known. With high expressive power of quantum circuits, this approach can be advantageous in various fields of science such as plasma physics and fluid dynamics where differential equations are ubiquitous.

\section{Importance of Co-design} 

Different quantum computing platforms have different strengths, weaknesses and peculiarities. Therefore, it is highly beneficial to customize quantum algorithms for a specific application that is suitable for the given architecture to obtain the maximum performance. In addition, co-design in quantum computing also pushes for hardware tailor made for some specific application~\cite{tomesh:codesign}. For example, variational algorithms, which rely on the continuous loops of quantum and classical computation, benefit from having the quantum and classical hardware be as close as possible in order to reduce the latency in information flow. Further, they require continuously tunable rotation gates which constitute the integral part of the variational ansatz. Hardware which naturally allow such continuously tunable and high fidelity gates are clearly more favorable for VQA applications. In addition, co-design has motivated researchers to use hardware-efficient variational ansatze which require little to no additional SWAP gates for the hardware implementation. This is a clear example of how NISQ devices can be exploited maximally for practical use with the help of co-design. Another promising direction towards quantum advantage in the near-term is to utilize them as accelerators in computation working alongside the CPUs using suitable classical-quantum hybrid algorithms. This would be analogous to the accelerating role of GPUs in the present cluster computing. An example of the power of co-design is observed in~\cite{hermanni:codesign}. Here the authors have resorted to the co-design of a quantum processor to implement the application specific task of simulating nanoscale NMR resonances. By coupling transmon qubits centrally to a co-planar waveguide resonator with a quantum circuit refrigerator, they have been able to achieve a 90\% reduction of SWAP operations for an all-to-all simulation of spins. As a consequence, the computation on the star topology was shown to reach the same accuracy of the computation with a 99\% two-qubit-gate fidelity that would require a 99.99\% two-qubit-gate fidelity on a square topology device. To reap the benefits of co-design, a highly cross-disciplinary collaborative effort among the researchers of hardware and software domains is necessary and constitutes an essential factor towards practical quantum advantage using NISQ devices.

\section{Challenges \& Opportunities} 

The biggest challenges of NISQ computing are that the available qubits (tens or hundreds) are imperfect and their operation is impacted by noisy processes, and the need for new and improved algorithmic ideas that can tolerate these limitations. The coherence times of qubits are low and therefore the success of quantum algorithms requires short depth circuits in such quantum computers, while making sure that non-classical correlations are present and operation is not classically simulable. Additionally, the limited qubit connectivity in the current NISQ device architectures impose overhead cost of qubit encoding and gate implementations. These contribute to the depth of the circuit. VQAs which are feasible on NISQ systems also face several challenges. During the parameter optimization, the derivative of the cost function can become vanishingly small because of “barren plateaus” in the parameter search space which can arise due to the random choice of ansatz, random initialization of ansatz ~\cite{Grant2019initialization} and the choice of cost function~\cite{cerezo_cost_2021}. 
Furthermore, quantum machine learning for classical problems requires efficient data encoding and initial state preparation and the measurement readout scales exponentially. These constitute the “input/output problem” of quantum machine learning which makes it difficult to compete with the existing classical algorithms. All these challenges put together point to the need for new ideas and more functional paradigms of NISQ algorithms. The hardware quality is steadily moving towards a regime where classical simulations of the best existing quantum computers is becoming extremely hard, if not impossible. Now we need to harness this capability with algorithms that use this capability for solving research or business problems. 

There is rigorous research being undertaken towards fault-tolerant quantum computers as well as to utilize the current NISQ devices for useful applications. Quantum Error Mitigation (QEM) is a technique designed for NISQ era quantum computing to reduce the noise effects as the existing quantum error correction (QEC) codes are not implementable on such devices. QEM techniques essentially implement several noisy circuits and the data obtained from these implementations is used to extrapolate the results in absence of noise.
There have also been many variants of the VQAs which overcome some of the drawbacks of their vanilla versions. An example is that of the Variational Fast-Forwarding (VFF) algorithm which allows the dynamic simulation of quantum systems for arbitrary time with fixed circuit length~\cite{cirstoiu_variational_2020} for a certain class of Hamiltonians that violate the time-energy uncertainty relation.


\section{Conclusions}

In this position paper, we have outlined the state-of-the art of NISQ devices, challenges, and their future prospects. Although it looks like a herculean task to build a full-fledged and fault tolerant quantum computer, we should not overlook the advancement that has been done so far. From the initial idea formulated by Feynman in the 80’s, Shor’s algorithm for factoring prime numbers with the superpolynomial speedup in the 90’s, at present we have working quantum computers which have shown quantum supremacy and are available to be remotely accessed by researchers for various experiments and further development~\cite{arute_quantum_2019}. 

While the race towards fault tolerant quantum computing continues, we strongly advocate for parallel research towards practical quantum advantages using NISQ devices. In this regard, particular effort needs to be dedicated to singling out a killer application; such is likely to be present within the field of quantum chemistry. These applications, which are seriously limited by the computational resources required to simulate them, can be the class of problems that might benefit from NISQ era computers. Merging AI with quantum computing techniques is another viable path towards achieving practical quantum advantage. While classical machine learning has already a plethora of applications, quantum machine learning techniques have the potential to yield substantial optimization given their enhanced expressivity. 

Another aspect concerns the necessity of software-hardware co-design for the NISQ era quantum computing. This promotes the fact that algorithm and hardware design need to be application specific and tailor-made, which is an essential component to utilize the NISQ devices to their maximum limit.  Hence, the crucial element in pushing the state-of-the-art would be to nurture collaborative efforts among physicists, computer scientists, material scientists and engineers. Given that there are plenty of computationally hard problems across various disciplines, a primary awareness of quantum computing capabilities among the experts of these fields is advantageous to identify potential problems that can benefit from quantum computing. The recent drives such as the \$5M global competition ``Quantum for real-world impact" by Google Quantum AI and XPRIZE~\cite{google-xprize}, and IBM Quantum's ``the $100 \times 100$ challenge"~\cite{ibm-100by100} are promising to accelerate the progress in employing current quantum computers for useful applications.
It is equally important to educate the public and investors with the state of the art and the realistic goals of near-term quantum devices without glamorizing or over-selling.

\section{Acknowledgments}
R.V. and S.M. acknowledge the KTH Digitalization Platform and Digital Futures for providing the funding to organize the workshop which motivated this work. P.R.H and S.M. acknowledges the funding from EuroHPC JU under the project Plasma-PEPSC, grant agreement No. 101093261. O.K. acknowledges the funding from UK EPSRC award under the Agreement EP/Y005090/1.
\bibliographystyle{ACM-Reference-Format}
\bibliography{sample-base}

\end{document}